# Harnessing the modulation instability spectrum in optical fibers with a periodic dispersion landscape

M. Droques, A. Kudlinski, G. Bouwmans, G. Martinelli and A. Mussot

*University of Lille 1, PhLAM Laboratory, IRCICA, 59655 Villeneuve d'Ascq, France*

*Corresponding author: alexandre.kudlinski@univ-lille1.fr

**Abstract:** We report the experimental demonstration of modulation instability process assisted by a dispersion grating in an optical fiber. A simple analytical model is developed to further analyze and explain the complex dynamics of this process, showing that each of the multiple spectral components grows thanks to a quasi phase-matching mechanism inherent to the periodicity of the waveguide parameters. This model is confirmed by numerical simulations and it is successfully used to tailor the multi-peak modulation instability spectrum shape. These theoretical predictions are confirmed by experiments.

**Keywords:** Dispersion management, Modulation instability, Nonlinear fiber optics.

**Introduction**

Modulation instability (MI) is an ubiquitous phenomenon in which a weak periodic perturbation of an intense stationary field is exponentially amplified. In optical fibers, it results from the interplay between chromatic dispersion and nonlinear Kerr effect. It is ruled by the nonlinear Schrödinger equation (NLSE) and it can also be interpreted in terms of four-wave mixing (FWM) formalism. For the amplification process to take place, it is thus required that the positive nonlinear phase mismatch (due to the focusing nature of the nonlinearity in glass optical fibers) is balanced by a negative phase mismatch originating from dispersion. In other words, dispersion must be anomalous in standard single-mode optical fibers. However, MI can still be observed in normally dispersive fibers providing that an additional degree of freedom is brought to the system. A common way consists in using birefringent fibers [1] or slightly multimode fibers [2]. Indeed, the required negative linear phase mismatch contribution can be obtained from the difference between the propagation constants of each involved mode. A more original way consists in adding another dimension to the system by periodically modulating the fiber properties along the propagation distance, which is the aim of our work.

Such a longitudinal periodicity can for instance be encountered in all optical telecommunication networks. Indeed, the alternation of regeneration (or dispersion compensation) optical devices and transmission lines creates a periodicity in power (or dispersion). The seminal work of F. Matera *et al*. [3] showed that the intrinsic periodicity of telecommunication networks may result in multiple new MI sidebands. Since then, a lot of studies have shown that MI may indeed be observed whatever the dispersion regime in periodic fiber systems [4-10]. In this specific context of optical telecommunication networks, the periodicity is thus highly detrimental because it leads to the generation of spurious multiple sideband pairs via FWM, which degrade the overall quality of the signal in wavelength-division multiplexing systems.

These studies however mainly remained of theoretical interest, and a clear experimental evidence of the presence of these multiple sidebands has still to be brought. Most experimental studies were performed either by cascading uniform fiber segments with different dispersion characteristics [11,12], or by periodically bending a uniform non birefringent fiber wrapped around two spools [13,14] or in a fiber cavity aiming at reproducing a telecommunication transmission line [15]. However, in all these experiments, only a single pair of sidebands was observed while the most streaking feature of the process lies in the multiple sidebands generation. In a recent experiment, modulated silicon waveguides were used to achieve broadband wavelength conversion [16], but again, thanks to a single frequency side band. Intriguingly, there has not been any experimental attempt to our knowledge to take benefit from the specific features of the MI process in optical fibers with continuous periodic dispersion while the fabrication of such optical waveguides was reported more than 20 years ago [17] and used for soliton fission management [18].

We report here for the first time to our knowledge the simultaneous generation of more than ten pairs of MI sidebands spaning over more than 10 THz in an optical fiber with a periodic dispersion landscape. We analyse these results and explain the complex nonlinear dynamics in terms of a quasi-phase-matching process with a simple analytical model. Finally, we take benefit from this analysis to design experiments demonstrating the possibility of tailoring the multi-peak MI spectrum thanks to the new degree of freedom brought by the periodicity.

**Materials and Methods**

Our experimental demonstration of MI assisted by a dispersion grating uses a photonic crystal fiber (PCF) which has been designed so that its group-velocity dispersion (GVD) is periodically modulated along the fiber axis, *z*. This can be done by periodically varying the fiber outer diameter during the fiber draw. Figure 1(a) shows the evolution of the outer diameter as a function of fiber length measured during the drawing process. It has a sinusoidal shape with a modulation period $Z$ of 10 m. The modulation amplitude corresponds to ±7 % of the average fiber diameter (117 µm) and the fiber length is 120 m (12 modulation periods). Figures 1(b) and (c) represent the scanning electronic microscope (SEM) images of the PCF cross sections recorded for extreme diameter values. They show that the ratio *d* (holes diameter) over Λ (pitch of the periodic cladding) is preserved along the fiber. From this observation, the longitudinal evolution of the GVD was calculated using the model from [19] and the d and Λ values extracted from SEM images. The longitudinal evolution of the second-order dispersion coefficient $\beta_2$ at the pump wavelength ($\lambda_P$ = 1072 nm) is plotted in Fig. 1(d). Since the relation between the outer diameter and $\beta_2$ is almost linear over the wavelength and outer diameter ranges of interest, $\beta_2(z)$ follows a quasi-sinusoidal shape with a positive average value ($\overline{\beta_2}$) of 1.2×10$^{-3}$ ps²/m and a modulation amplitude ($\beta_2^A$) of 1.5×10$^{-3}$ ps²/m (i.e. a modulation rate of ±125 %). The corresponding zero-dispersion wavelength (ZDW) varies between 1069 nm and 1097 nm. The relative longitudinal variations of the NL coefficient $\gamma$ are much lower (±12 %), so that the fiber can indeed be seen as a mainly dispersion-managed device, which is the reason why we named it dispersion oscillating fiber (DOF).

**Results and Discussion**

This DOF (labeled DOF#1) was pumped by a tunable fiber master optical power oscillator delivering linearly polarized 2 ns square-shaped pulses. Figure 2(a) illustrates a typical multi-peak spectrum observed at the DOF output. In this particular example obtained for $\lambda_P$ = 1072 nm and a pump power $P_P$ of 20 W, more than ten spectral components are generated on both sides of the pump. They are not equally spaced, which indicates that these spectral components are not harmonics of the first side lobe, conversely to the standard cascaded FWM process in uniform optical fibers [20]. Numerical results displayed in Fig. 2(b) are based on the numerical integration of the generalized NLSE [21]. Numerical results obtained in these conditions confirm our experimental observations, as shown by the good agreement between Figs. 2(a) and (b). Slight discrepancies between both spectra may be attributed to small uncertainties on experimental parameters (fiber characteristics, pump power, ...), and to the high experimental noise floor caused by residual amplifies spontaneous emission of the pump. It is worth noting that in the present configuration, the pump at 1072 nm experiences a normal dispersion over the whole fiber length at the exception of very short regions located around the minimum of the GVD curve, as displayed in Fig. 1(d). Thus the average GVD at the pump wavelength is largely normal so that, without any periodic GVD modulation, absolutely no spectral component would be generated (in the case of a single-mode scalar propagation and neglecting higher-order dispersion). By the way, we voluntary choose this specific dispersion region because it allows to the generation of the highest number of sidebands over a broad spectrum. It is of course possible to work in a pure normal dispersion region but one would not obtain such outstanding results.

To confirm that these multiple sidebands have a parametric origin, we investigated the gain properties of such DOFs. For practical reasons, a slightly different sample (labeled DOF#2) from the one displayed in Fig. 3 was used for this measurement. DOF#2 is 120 m-long and it has an average dispersion $\overline{\beta_2}$ of 0.8×10$^{-3}$ ps²/m, a modulation amplitude $\beta_2^A$ of 10$^{-3}$ ps²/m at 1067 nm and its ZDW oscillates between 1064 nm and 1080 nm. The average third and fourth order dispersion terms are respectively $\beta_3$ = 6.8×10$^{-41}$ s³/m and $\beta_4$ = 1.7×10$^{-55}$ s⁴/m, the average nonlinear coefficient is $\gamma$ = 7 W$^{-1}$.km$^{-1}$ and the attenuation is $\alpha$ = 7.5 dB/km. The pump was the same as described above, but it was tuned to a wavelength of 1067 nm and a peak power of 26 W. The signal used to seed the amplifier was a single-mode cw tunable laser diode with an average power of 10 µW. Its polarization was linear and aligned to that of the pump thanks to a polarization maintaining coupler before being launched inside the fiber along one of its polarisation axis. As the pump is pulsed and the signal is cw, the gain terminology used in the following refers to the peak gain, i.e. the gain experienced by the signal during a pump pulse. Crosses in Fig. 3 shows the measured parametric peak gain spectrum over the only first three sidebands because we were limited by the tunability of our probe source to go beyond. We obtained about 50 dB of amplification on these three lobes, illustrating the strong efficiency of this parametric process. These results were confirmed by a numerical integration of the generalized NLSE, represented in solid black line in Fig. 3. The MI process in DOFs thus allows to obtain high gain values over multiple bands with large detunings from the pump (up to about 10 THz). It is important to make a clear difference between the FWM process investigated here and so-called Kelly sidebands which have been experimentally observed in the early years of soliton fiber lasers [22,23]. These processes are by nature different [Matera1993] since Kelly sidebands originate from the destabilization of solitons in active fiber cavities [24,25], while the process investigated here is parametric as proven the gain measurement experiment.

The frequency of these parametric sidebands can be estimated by a quasi-phase-matching relation developed to characterize this process for an infinitely long grating [3-5]:

$$\overline{\beta_2}\Omega_k^2 + 2\gamma P_P = \frac{2\pi k}{Z} \quad (1)$$

where $k$ is integer and $\Omega_k$ is the pulsation detuning from the pump. This simple quasi-phase-matching relation predicts rapidly and quite accurately the frequency of amplified spectral components (see arrows in Fig. 2). The discrepancy between theoretical values from Eq. 1 on one hand and experimental/numerical values on the other hand is indeed less than 15% for $k$ = 1 to 10, which is partially due to fiber losses, which were neglected to obtain Eq. 1, and to a more fundamental reason that will be discussed hereafter.

In order to investigate the spectral dynamics of this process, the evolution of the spectrum over the length of DOF#1 was experimentally recorded (by sequentially cutting back the fiber) and numerically studied. The results are plotted respectively in Figs. 4(a) and (b). All side lobes start to grow from the beginning of the DOF, as in a standard phase-matched FWM process, but due to the experimental noise floor, the measurement was only possible from 40 m. The dynamics reveals that the amplification process is not monotonic over the whole fiber length and small oscillations of the power and maximal gain frequency can be observed for all spectral components both in experiments and simulations. For the sake of simplicity, we focus our attention on the first side lobe ($k$ = 1). First, the inset of Fig. 4(c) shows that the frequency of maximal gain clearly oscillates within the first periods, and it tends to the quasi-phase-matched pulsation $\Omega_k$ predicted from Eq. 1 [displayed by the horizontal dashed line in Fig. 4(c)] for a long propagation distance. This behavior may be interpreted considering the evolution of the accumulated linear phase mismatch during the propagation. This phase value tends to $\overline{\beta_2}\Omega^2\ell$ for $\ell/Z \gg 1$, with $\ell$ the DOF length, but for shorter $\ell$ values, it oscillates around this mean value with an amplitude of modulation that decreases for increasing distance. Thus the maximum gain frequency tends to the quasi-phase-matched frequency given by Eq. 1 in the long propagation distance regime ($\ell/Z \gg 1$) while it oscillates around this frequency for shorter $\ell$. Second, the longitudinal evolution of the maximum power of the first side lobe is plotted in Fig. 4(c) for experiments (crosses) and simulations (solid line). There is a very good agreement between experimental and numerical results except at short fiber lengths, which could be due to the rather poor experimental signal to noise ratio in this domain. The overall average gain is exponential (as expected for a parametric process with negligible pump depletion) but the dynamics study shows that it oscillates around this general trend at the same period than the one of the fiber dispersion map (10 m). This is similar to quasi-phase-matching processes in $\chi^{(2)}$ crystals in which the nonlinear characteristics of the medium are periodically tailored to compensate for the phase-mismatch between all waves along propagation. However, the dynamics observed in the DOF exhibits periodic regions of desamplification, which makes the side lobe power oscillate along the fiber around the exponential growth (expected for a perfectly phase-matched FWM process), as mentioned above. This particular and unusual feature will now be studied in detail with help of a simple analytical model, in order to provide further insight into the underlying physics.

The parametric gain spectrum of a FWM process can be obtained by studying the stability of the steady state solution against weak perturbations through a so-called linear stability analysis. In DOFs, this tool has allowed to analytically predict the complex multi-peak gain spectrum [Smith1996, Abdullaev1996], but such an analysis does not provide a clear insight into the dynamics of the process. To this aim, we propose here a more intuitive explanation of the results above by revisiting a simplified three wave model usually aimed at describing Fermi Plasta Ulam recurrence and fiber-optic parametric amplification [26-28]. This model allows to account for the relative phase variations between the pump, the signal and the idler waves during the propagation. In our work, it will be induced by the longitudinal variations of the dispersion rather to the pump depletion. In order to simplify our investigations, our starting point is the four coupled differential equations given by Eqs. 3 in Ref. [26]. We neglect fiber loss, we assume that the pump remains undepleted, that signal and idler powers, $P_S$ and $P_I$, are much less than the pump power $P_P$ over the whole DOF length. By additionally including the longitudinal variations of dispersion, this system reduces to the following equations :

$$\frac{dP_S(\Omega, z)}{dz} = 2\gamma P_P \sqrt{P_S(\Omega, z) P_I(\Omega, z)} \sin \theta(\Omega, z) \qquad (2)$$

$$\frac{dP_I(\Omega, z)}{dz} = 2\gamma P_P \sqrt{P_I(\Omega, z) P_S(\Omega, z)} \sin \theta(\Omega, z) \qquad (3)$$

$$\frac{d\theta(\Omega,z)}{dz} = \Omega^2\left[\overline{\beta_2} + \beta_2^A \sin\left(\frac{2\pi z}{Z}\right)\right] + 2\gamma P_P\left[1 + \cos\theta(\Omega,z)\right] \qquad (4)$$

where Ω is the shift of the signal and idler pulsations from the pump, and $\theta(\Omega,z)$ describes the longitudinal evolution of the relative phase difference between all these waves [26]. The above mentioned discrepancy between solutions of Eq. 1 and experimental/numerical values can now be understood from Eq. 4. Indeed, Eq. 1 assumes that the nonlinear phase mismatch can be approximated by $2\gamma P_P$ [3-5], while Eq. 4 shows that it is in fact equal to $2\gamma P_P[1+\cos\theta(\Omega,z)]$ in an axially varying fiber. This important point will be further detailed in an upcoming paper, but it does not impact the validity of the present results since this additional term remain low for the pump powers involved in the present study. In order to obtain a simple analytic solution of the set of Eqs. 2, 3 and 4, we thus neglect the last term $\cos\theta(\Omega,z)$ in Eq. 4. It physically means that we assume that the longitudinal evolution of the nonlinear phase mismatch term is weak as compared to the linear and uniform nonlinear phase mismatch terms, which is valid for low pump powers. By using a Fourier series decomposition, we found that the signal gain in power writes as :

$$G(\Omega,z) = \frac{P_S(\Omega,z)}{P_S(\Omega,0)} = \frac{1}{4}(1-\rho) + \frac{1}{4}\left(1+\rho+2\sqrt{\rho}\right)\exp\left[\int_0^z g(\Omega,z')dz'\right] \qquad (5)$$

with $\rho = P_I(\Omega,0)/P_S(\Omega,0) = 1$ in all this study, and

$$g(\Omega,z) = 2\gamma P_P \sum_{q=-\infty}^{q=+\infty} J_q\left(\frac{\beta_2^A \Omega^2}{2\pi/Z}\right)\sin\left[\left(\overline{\beta_2}\Omega^2 + 2\gamma P_P - \frac{q 2\pi}{Z}\right)z + K_q\right] \qquad (6)$$

with $K_q = \frac{\beta_2^A \Omega^2}{2\pi/Z} - q\frac{\pi}{2} + \theta(\Omega,0)$. Thus, Eq. 6 indicates that the linear gain $g(\Omega,z)$ at a fixed pulsation detuning Ω can be interpreted as the sum of sinusoidal functions in z. These sinusoidal functions all have a zero average value except when their argument becomes independent of z. It occurs only at specific spectral components $\Omega_q$ corresponding to solutions of the quasi-phase-matching relation 1. For these specific pulsation detunings $\Omega_q$, each term of the sum in Eq. 6 leads to periodical amplification and desamplification phases along the DOF except for the uniform contribution corresponding to q = k. This last term therefore prevails over the other ones on the gain $G(\Omega,z)$ for long enough propagation distances. Thus the linear gain of the $k^{th}$ spectral component can be approximated by this uniform term $2\gamma P_P \left|J_{q=k}\left(\frac{\beta_2^A \Omega_{q=k}^2}{2\pi/Z}\right)\right|$, with

$K_q = +\frac{\pi}{2}$ in order to maximize the uniform gain by adjusting $\theta(\Omega_q, 0)$. This is physically analogue to the self-adjustment of the idler phase value at the beginning of the fiber in a phase insensitive parametric amplifier [26].

To illustrate this, we focus on the first amplified spectral component (*k* = 1). The solid black line in Fig. 5 shows the evolution of the maximum gain (at $\Omega = \Omega_{\max}^{simu}$) obtained from numerical integration of the complete set of original equations (Eqs. 2, 3 and 4). Note that an excellent agreement is achieved with the numerical integration of the GNLSE (not shown here for the sake of clarity). The red dashed line in Fig. 5 corresponds to the term of uniform gain (Bessel function $J_{q=1}$), the red dotted line corresponds to the highest amplitude oscillating term (Bessel function $J_{q=0}$ in this case) and the red solid line corresponds to their sum. We limit our investigations to $J_{q=0}$ because all other Bessel functions have much lower contributions in this example. A good agreement is obtained between the red solid curve from the analytical model, and the black one from numerical simulations, which confirms the validity of our assumptions and the accuracy of our method. In each modulation period, the amplification phase is characterized by $0 < \theta(\Omega, z) < \pi$ and the desamplification one has $-\pi < \theta(\Omega, z) < 0$, the total phase shift being equal to 2π per period (2*k*π for the *k*th spectral component). The dispersion modulation therefore allows to control the evolution of the relative phase of the waves and the whole process can thus be seen as a quasi-phase-matching. More precisely, the frequency of the spectral component Ω*k* can be widely modified simply by changing the periodicity of the grating (as in a diffraction grating for the position of its different orders), while the gain (analogue to the diffraction efficiency in a specific order) can be modulated independently through the ratio $\beta_2^A / \overline{\beta_2}$. Note however that the desamplification phases of the signal along propagation cannot be totally avoided, since they are due to the contribution of all oscillating terms of Eq. 6 which cannot be all suppressed simultaneously.

This simple analytical approach allows a better understanding of the complex dynamics of the process and it has allowed us to design experiments in which the multi-peak gain spectrum is tailored. To illustrate this, we focus our attention here to two streaking examples. We chose either to completely cancel a given spectral component or to maximize a sideband pair with regards to the others. To reach these goals, let us recall that, as detailed in the analytical model above, the linear gain of the kth spectral component can be approximated by:

$$g(\Omega_k, z) = 2\gamma P_P \left| J_{q=k}\left(\frac{\beta_2^A \Omega_k^2}{2\pi/Z}\right) \right| = 2\gamma P_P \left| J_{q=k}\left[\frac{\beta_2^A}{\overline{\beta_2}}\left(k - \frac{\gamma P_P Z}{\pi}\right)\right] \right| \qquad (7)$$

Equation 7 indicates that the gain of the $k^{th}$ spectral component can be totally cancelled by simply finding the argument $\eta = \dfrac{\beta_2^A \Omega_k^2}{2\pi/Z} = \dfrac{\beta_2^A}{\overline{\beta_2}}\left(k - \dfrac{\gamma P_P Z}{\pi}\right)$ for which the Bessel function $J_{q=k}$ vanishes. $\eta$ can be adjusted by controlling the modulation amplitude of dispersion ($\beta_2^A$) or the fiber period ($Z$) which both requires manufacturing new DOF samples. But it can also be adjusted by controlling the average dispersion at the pump ($\overline{\beta_2}$). This can be done experimentally by simply tuning the pump wavelength. To illustrate this, we used DOF#2 [dof] and, for example, choose to cancel the $k = 6$ sideband pair. In this case, we found that $J_6$ vanishes for a $\overline{\beta_2}$ of 5.8×10$^{-4}$ ps²/m with the parameters of DOF#2 given above and a pump power of 13 W. Squares in Fig. 6(a) shows the gain calculated with the above model (Eq. 7) for each spectral component and for a fiber length of 120 m, while the solid line represent the output spectrum obtained from a numerical integration of the generalized NLSE with a pump power of 13 W. We average 50 output spectra seeded by random initial conditions to account for the averaging performed during the experimental recording of a spectrum. These results firstly confirms the ability of our simplified model to correctly predict the maximal gain of each sideband and they also show that the $k = 6$ spectral component is indeed cancelled. Experiments performed in DOF#2 by tuning the pump wavelength to 1067.5 nm (which is close to the required $\overline{\beta_2}$ value of 5.8×10$^{-4}$ ps²/m) are displayed in Fig. 6(b). The overall shape of the experimental spectrum nicely matches the one obtained from theory and this measurement also confirms the cancellation of the 6$^{th}$ peak, which reinforces the validity of our analytical approach. In all experiments presented in this section, the pump power was the only adjustable parameter. It had to be adjusted to 24 W to observe the expected behaviors, which is slightly higher than the power of 13 W used in simulations and in the model. In order to further illustrate the possibility of manipulating the multipeak MI spectrum, we used Eq. 7 to find a configuration in which the $k = 1$ sideband is maximized, i.e. it experiences a much higher gain than any other sidebands. In this case, we simply need to find a $\overline{\beta_2}$ value (and thus a value of the $\eta$ parameter) which maximizes the $J_1$ Bessel function. Figure 6(c) shows the gain calculated from Eq. 7 in squares, as well as the output spectrum obtained from numerical integration of the generalized NLSE for a $\overline{\beta_2}$ value of 3.87×10$^{-4}$ ps²/m. These results are again in excellent agreement, and they indeed show that the first sidelobe is favored since it has a 25 dB gain higher than all other ones. Experiments were performed by accordingly tuning the pump wavelength to 1071.5 nm. The output spectrum plotted in Fig. 6(d) shows that the power of the first sidelobe is 22 dB higher than other spectral components, in good agreement with theoretical predictions.

Although the control of the overall MI spectrum shape requires a change of quasi-phase-matched frequencies, these examples demonstrate the possibility of harnessing the MI spectrum thanks to the periodic dispersion landscape. A simultaneous control of both the spectral shape and sideband frequencies would still be possible by simultaneously adjusting $\overline{\beta_2}$ and $\beta_2^A$, which would however require manufacturing new fibers.

**Conclusion**

We have experimentally reported for the first time the generation of multiple sideband pairs over a wide spectrum through MI in an optical fiber with periodic dispersion. The process exhibits a complex dynamics showing periodic oscillations of the maximum gain and spectral variations of the generated components. A simple model has been developed to interpret these results in terms of a quasi-phase-matching process inherent to the periodic nature of the dispersion landscape. This model has been used to predict the possibility of tailoring the MI spectrum, which has been confirmed experimentally by the cancellation or maximization of chosen spectral components.

The new degree of liberty brought by the periodic dispersion landscape offers a very innovative platform for fundamental studies of nonlinear dynamics, such as the development of instabilities, as reported here. From a practical point of view, these potential wide gain bands should lead to important breakthroughs in short pulse generation and/or amplification.

**Acknowledgments**

We acknowledge J. M. Dudley for fruitful discussions, I. Dahman for the gain measurement and K. Delplace for assistance in fiber fabrication. This work was partly supported by the French Ministry of Higher Education and Research, the Nord-Pas de Calais Regional Council and FEDER through the "Contrat de Projets Etat Région (CPER) 2007-2013" and the "Campus Intelligence Ambiante" (CIA).

**List of figures**

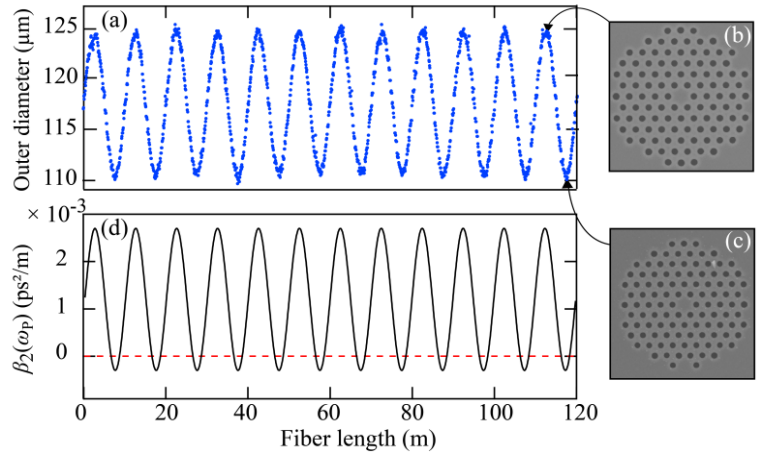

Fig. 1: (a) Evolution of the outer diameter as a function of fiber length measured during drawing. (b)-(c) SEM images (at the same scale) of the fiber cross-section at respectively the maximum and minimum diameters. (d) Calculated evolution of the second-order dispersion coefficient $\beta_2$ at the pump wavelength $\lambda_P$ = 1072 nm deduced from SEM images. The dashed line represents the pump wavelength.

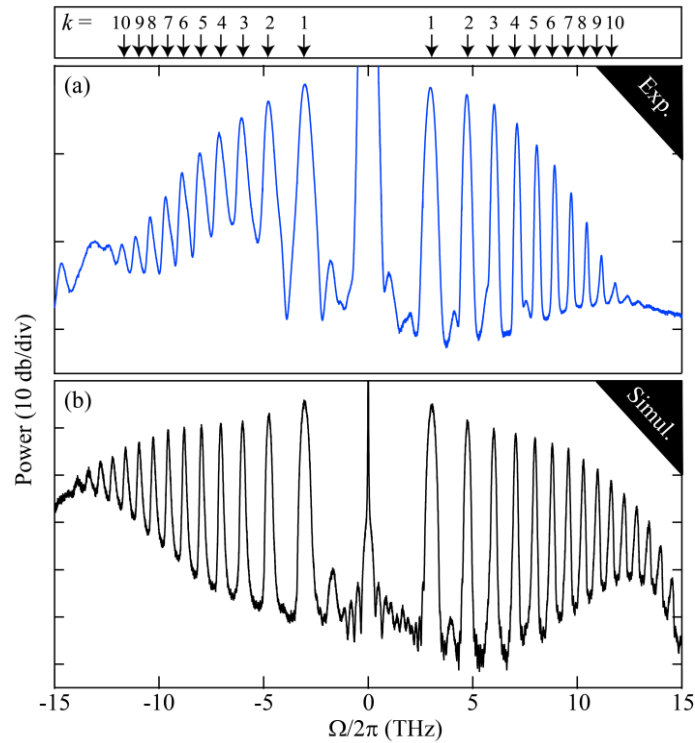

Fig. 2: (a) Experimental and (b) numerical spectra out of the 120 m long DOF#1 for the same pump power of 20 W. Note the different scale due to a lower dynamics in experiments compared to simulations. Arrows at the top represents quasi-phase-matched frequencies $\Omega_k$ obtained by solving Eq. 1 for $k$ = 1 to 10.

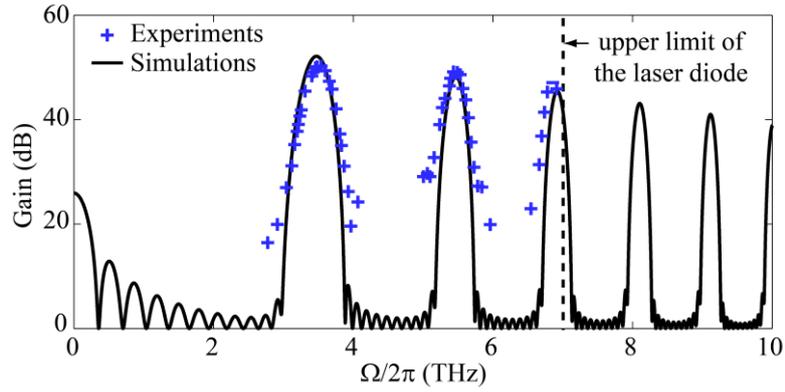

Fig. 3: Amplification peak gain in DOF#2. Experimental results are depicted by crosses and numerical simulations appear in solid black line. The vertical dashed line represents the upper tunability limit of the laser diode used for the signal.

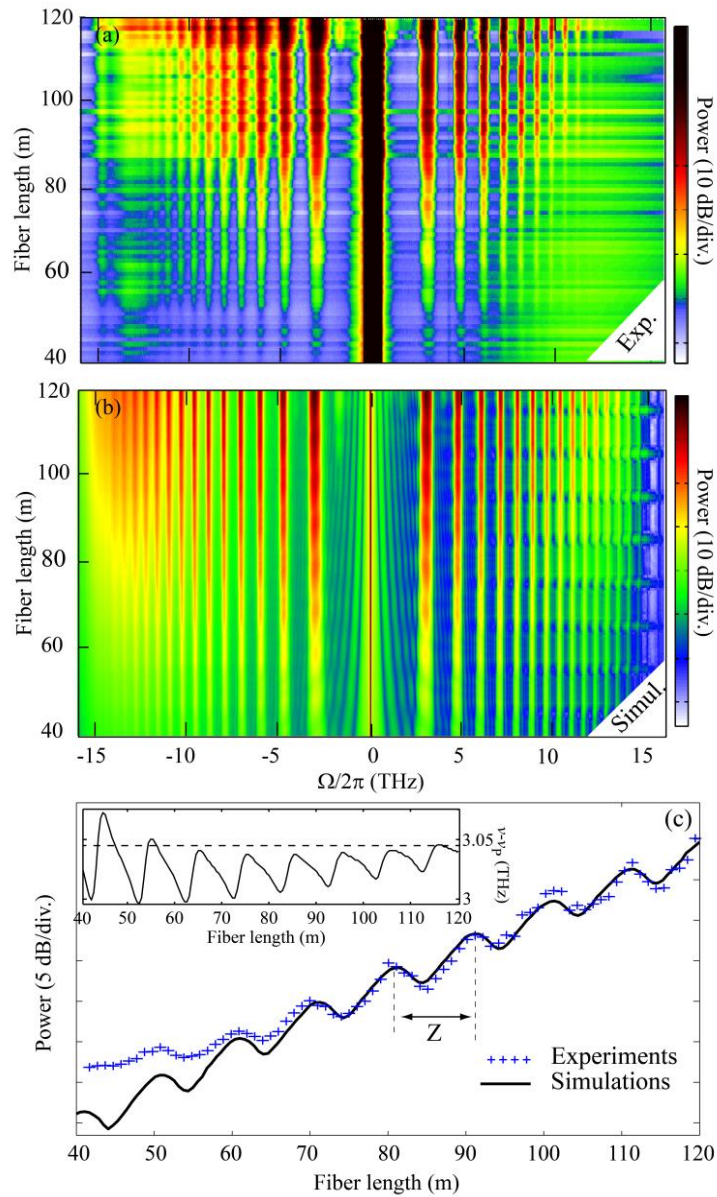

Fig. 4: (a) Experimental and (b) numerical dynamics of the structured spectrum formation in the DOF. (c) Evolution of the power of the first side lobe obtained from experiments (crosses) and numerical simulations (solid line). Inset: evolution of the maximum gain frequency obtained from numerical simulations.

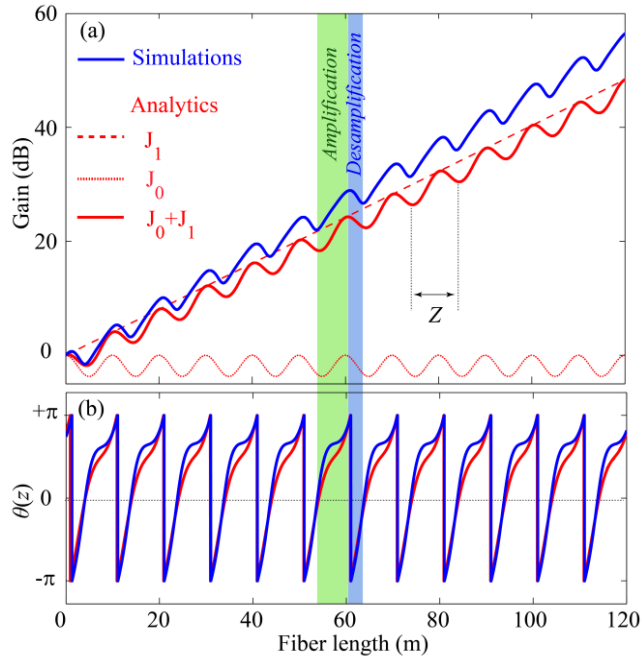

Fig. 5: (a) Evolution of the gain of the first amplified frequency ($k = 1$) from Eq. 5, in red lines with the contribution of $J_1+J_0$ (average gain + oscillating term), in red dashed lines for $J_1$ only and in red dotted lines for $J_0$ only with $\Omega_{max}^{theo}=2\pi\times2.63\times10^{12}$ rad/s (Eq. 1). The solid black line is calculated from the numerical integration of the original set of equations (Eqs. (3) in Ref. [26]) with $\Omega_{max}^{simu}=2\pi\times2.93\times10^{12}$ rad/s. (b) Evolution of $\theta(\Omega)$ from our analytical study (solid red line) and from the numerics (solid black line).

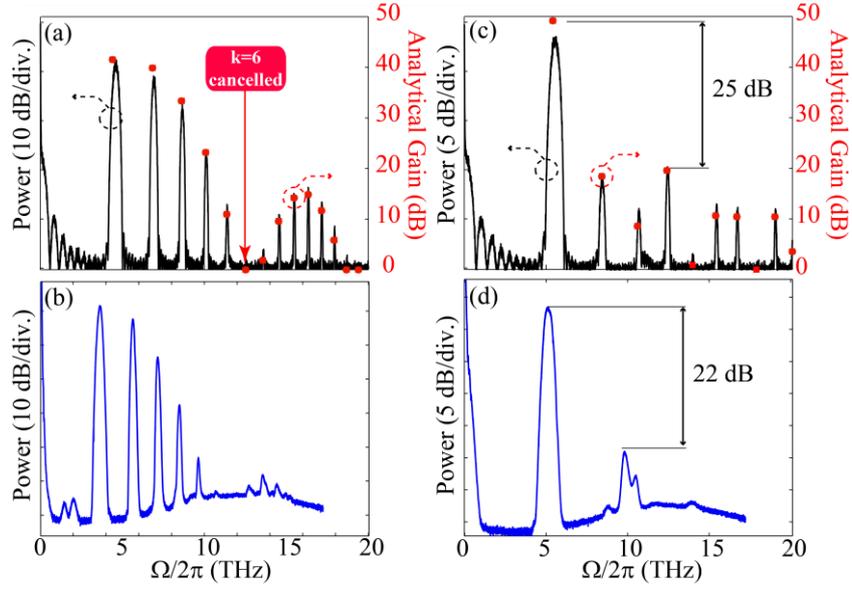

Fig. 6: (a)-(b): Illustration of the cancellation of the 6th spectral component. (a) Maximal gain obtained from Eq. 7 (squares, right axis) and output spectrum simulated with the generalized NLSE (solid line, left axis), for $\overline{\beta_2}$ = 5.8×10⁻⁴ ps²/m) and $P_P$ = 13 W. (b) Corresponding experiments performed in DOF#2 for a pump wavelength of 1067.5 nm and pump power of 24 W. (c)-(d): Illustration of the maximization of the 1st spectral component. (c) Maximal gain obtained from Eq. 7 (squares, right axis) and simulated output spectrum (solid line, left axis), for $\overline{\beta_2}$ = 3.87×10⁻⁴ ps²/m and $P_P$ = 13 W. (d) Corresponding experiments performed in DOF#2 for a pump wavelength of 1071.5 nm and pump power of 24 W.